# Perception of corruption influences entrepreneurship inside established companies

**F. Javier Sánchez-Vidal**
Dpto. de Economía, Contabilidad y Finanzas
Universidad Politécnica de Cartagena
Email: javier.sanchez@upct.es

**M. Camino Ramón-Llorens**
Dpto. de Economía, Contabilidad y Finanzas
Universidad Politécnica de Cartagena
Email: camino.ramon@upct.es

**ABSTRACT**

Based on the Global Entrepreneurship Monitor (GEM) surveys and conducting a panel data estimation to test our hypothesis, this paper examines whether corruption perceptions might sand or grease the wheels for entrepreneurship inside companies or intrapreneurship in a sample of 92 countries for the period 2012–2019. Our results find that the corruption perception sands the wheel for intrapreneurship. There is evidence of a quadratic relation, but this relation is only clear for the less developed countries, which sort of moderate the very negative effect of corruption for these countries. The results also confirm that corruption influences differently on intrapreneurship depending on the level of development of the country.

1. **Introduction**

Entrepreneurial activity is one of the most determining engines of economic growth and development of a country (Gil-Soto et al., 2017; Almodóvar-Gonzalez et al., 2019) and a topic of growing interest to policymakers who strive to promote such entrepreneurial activity (Ács & Szerb, 2007). Entrepreneurship is a multi-dimensional concept (Yamada, 2004) that has been the object of great consideration in the academic research. Moreover, in last years it has been pointed out that entrepreneurship need not be filled by independent entrepreneurs and does not have to refer only to the creation of new companies, but can also be carried out within an established organization by entrepreneurial employees, also known as intrapreneurs (Pandey et al., 2020; Stam, 2019). In this regard, as a company matures, growth rates grow increase less and less and may even decline. At that time the company must strive to survive, seeking higher growth rates through innovation and creativity (Veronica et al., 2013) by taking advantage of the favorable conditions available in the external environment (Yoruk, 2019). It is here precisely where the concept of intrapreneurship arises.

Previous literature indicates that entrepreneurship and its effects are unevenly distributed geographically (Beynon et al., 2016, Beynon et al., 2018), and it depends on social, political, and cultural factors able to impact the development of an entrepreneurial environment.

Corruption, defined as misuse of public power for private benefit (Rodriguez et al., 2006) is considered as an important characteristic of a country's institutional quality (Dreher & Gassebner, 2013) that influences on entrepreneurial and innovative activity across nations. When a country is faced to inefficient institutions, such as mafia and corruption, entrepreneurship activities could be affected (Douhan and Henrekson, 2010) due to the fact that it misrepresents the individual perception of the governance capacity, which falls in inefficiency due to a bureaucratic governance structure (Méon & Sekkat, 2005).

Literature has studied the connection between corruption and entrepreneurship with inconclusive results. Evidence suggests that the relationship between these two variables depends on contextual factors, both at an individual and national level of analysis (Uribe-Toril et al., 2019). On the one hand, the perception of corruption might have a sand-the-wheel effect on entrepreneurial activities since it creates a climate of uncertainty towards government bodies, which can discourage entrepreneurial intentions. Researchers who support of this hypothesis consider that corruption cannot, in any case, have positive effects on business activity (Mauro 1995; Dutta & Sobel, 2016; Xu & Yano, 2017). However, on the other hand, the perception of corruption may grease-the-wheel of entrepreneurship when it allows entrepreneurs to overcome the difficulties induced by institutional dimensions (North, 2005; Ceresia and Mendola, 2019). When companies face to inefficient governments that waste resources or incur in bribery and other corrupt practices, or when they are overcome by excessive, inflexible and not transparent regulations, corruption acts as an economic catalyzer, facilitating the required administrative procedures for the entrepreneur, since obtaining credits and licenses from the authorities through legal channels can become a difficult task (Hanoteau and Vial, 2014).

Based on the Global Entrepreneurship Monitor (GEM) 2011 database, this study goes further and analyze how the intrapreneurs activity can be affected by the perception of corruption across the globe. Our sample includes the data for intrapreneurship, the perception of corruption for an 8-year period and for 92 countries with different development levels. Our findings indicate that corruption exerts a negative influence on intrapreneurship. There is evidence of a quadratic relation between corruption and intrapreneurship and this evidence is significant for the less developed countries.

This paper makes several contributions to the field. First, to the best of our knowledge is the first paper that provides an insight into the relation of intrapreneurship and perception of corruption. The second thing is that from a methodological point of view it takes into account the quadratic relation and the strong individual effects that entrepreneurship in general and intrapreneurship specifically show, and that determine the analyses. The rest of this study is designed as follows. Section 2 includes a brief literature review as a base to develop our hypothesis. Section 3 discusses the data and methodology used to test our hypothesis. Section 4 offers the results and their discussion and finally the conclusions are considered in Section 5.

## 2. Literature Review and Research Hypotheses

*2.1. Corporate entrepreneurship*

According to the data provided by Global Entrepreneurship Monitor (GEM) it is possible to distinguish independent entrepreneurial activity by individuals beginning and managing a business for their own account and risk, and corporate entrepreneurship, also named entrepreneurial employee activity (EEA) or intrapreneurship[1], which consists of either creating a new company within an existing organization, or to take advantage of innovative activities and orientations such as the development of new products or services, technologies and strategies, with the aim of creating economic value (Guerrero and Peña Legazkue 2013; Antoncic and Hisrich, 2003). Intrapreneurship is "entrepreneurship inside of the corporation" (Pinchott, 1985). Intrapreneurship can be a value-creating strategy, since it implies diversification, either related (Markides and Williamson, 1996) or unrelated (La Rocca et al., 2018)

There are three different levels through which we can analyze intrapreneurship: at the individual level, at the firm level and at the level of the aggregate economy (Stam, 2019). At the individual micro level, human capital literature can explain why a person chooses to be an intrapreneur. In this regard, the intrapreneur's profile such as the level of education, experience and knowledge (Davidsson and Honig, 2003), age (Levesque and Minniti, 2006), gender (Leoni and Falk, 2008), marital status (Özcan, 2011), as well as family background (Dunn and Holtz-Eakin, 2000) are factors influencing the probability of becoming an intrapreneur. At a firm micro level, intrapreneurship, also known as corporate entrepreneurship (Bosma & Guerrero, 2013) can be reinforced by some organizational factors such as achievement spirit, search for opportunities, risk assumption, proactiveness and innovativeness (Rigtering and Weitzel, 2013; Camelo-Ordaz et al. 2012; Jong Jeroen de, 2017). Besides, the behavior of companies within the same region are likely to be related (Maté et al, 2013), and so, the influence of corruption in a company's intrapreneurship could also be contagious, increasing its impact. Finally, there are some macro-level factors which influence the intrapreneurship activity, such as the cultural and social development of a country, (Klofsten et al., 2020; Honig and Samuelsson, 2020) and the social climate of the society, which includes sociological, economic, and institutional factors (Schumpeter, 1934).

In this regard, intrapreneurship might be regarded as critical for welfare in advanced economies (Stam, 2013), exerting a clear positive effect for society (Neessen et al., 2019). From this perspective, the Global Monitor Entrepreneurship (GEM) confirms that those countries with higher rates of intrapreneurship generate more jobs and are more competitive than those countries that, on the contrary, have more "conventional" entrepreneurs.

---

[1] Although both concepts are related (intrapreneurship *vs* corporate entrepreneurship) they are not exactly the same (Antoncic & Hisrich, 2003; Sharma & Chrisman, 1999). Corporate entrepreneurship is a top-down process that is defined at the company level and refers to management strategies to foster workforce initiatives and efforts to innovate and develop new businesses. Intrapreneurship refers to those bottom-up and proactive initiatives related to the work of individual employees. The way our dependent variable is defined (see section 3.1) involves these two types of related notions.

According to the institutional theory, institutional determinants in a society such as the social, political, and cultural frameworks, shape the structural attributes or "the rules of the game" within which entrepreneurs operate (North 1994, p. 361). These institutional determinants can be either formal or informal. The former are represented by the political, legal and economic system, rule and regulations, procedures, etc., which are erected by the governance of a nation to control the behavior of individuals within it (Dheer, 2017). Informal institutions on the other hand are composed of culture, norms, customs, values, beliefs, attitudes as well as the perceptions of the policies and regulations implemented by governments (North 2005; Sánchez-Vidal et al. 2012; Stenholm et al., 2013).

Previous research shows that informal institutions have a greater impact on entrepreneurship than formal ones (Urbano and Alvarez, 2014) indicating a positive relationship between favorable governance indicators and entrepreneurial activity (Aidis et al., 2008). When a country is faced to inefficient institutions, associated with mafia and corruption, entrepreneurship activities could be affected (Douhan and Henrekson, 2010), as it distorts the individual perception of the governance capacity, which falls in inefficiency due to a bureaucratic governance structure (Méon and Sekkat, 2005*)*. Entrepreneurship development can be adversely affected in those countries with higher levels of corruption (Akimova, 2002). Policy makers can play an active role in promoting growth by creating an appropriate environment for firms and/or encouraging certain policies (Sanchez Vidal et al, 2020).

*2.2. Corruption as an influencing factor in intrapreneurial activity*

Previous research has analyzed the effects of corruption on entrepreneurship with conflicting results. On the one hand, a stream of literature seems to confirm the "sand-the-wheels" hypothesis. The perception of corruption negatively impacts the entrepreneurial development and economic growth of a country intention due to the feeling of mistrust that it provokes and the idea that the business will be unsuccessful (Allini et al., 2017; Xu & Yano, 2017; Dutta and Sobel, 2016; Del Mar Salinas-Jiménez and Salinas-Jiménez, 2007). On the other hand, alternative research supports the

"grease-the-wheels" hypothesis which states that corruption can serve as an economic catalyst by encouraging the creation of new companies in those countries where it is more difficult to carry out business activity due to the numerous barriers they must face, such as strict rules and regulations or credit restrictions, among others. Corruption generate a grease the wheel effect as it helps entrepreneurs overcome bureaucratic limitations as is able to simplify business procedures and processes (Dreher & Gassebner, 2013). Entrepreneurs who work in a corrupted environment not only tend to integrate it into their ordinary activities but also end up interpreting it as a feasible and acceptable practice as it allows them not only to reduce uncertainty and business risks (Djankov et al., 2005; Harbi & Anderson, 2010; Ceresia & Mendola, 2019) but also to overcome the difficulties caused by institutional dimensions (North, 1991 & 2005; Urbano & Alvarez, 2014). Logically, this arguable behavior will depend on each country's culture and on the managers' personality, and for example personality traits have been found to play a role in some firm's policies, which could be considered unethical or at least debatable (García-Meca et al. 2021).

In the light of this, we hypothesize that:

*Hypothesis 1. The level of corruption perception affects the intrapreneurship activity*

Whereas both hypotheses (grease or sand) seem contradictory, they could coexist within the same entrepreneurial ecosystem (Mauro, 1995). This is because there are certain contextual factors which can provoke corruption to affect company's bottom line differently. In this regard, evidence suggests that a better control of corruption in developed countries contributes to increase the entrepreneurial activity (Wennekers et al., 2005). Regarding developing countries, corruption discourages entrepreneurship activities through the large barriers they face to (Bohata and Mladek, 1999; Johnson et al., 2000). Going back to entrepreneurship, it is a fact that is not the same across nations. This can be explained mainly by these macro-level factors (Liñán and Fernandez-Serrano 2014; Urbano and Alvarez, 2014). A similar thing can be expected to happen at an intrapreneurship level. Our second hypothesis is based on these arguments:

*Hypothesis 2. The level of corruption perception affects the intrapreneurship activity differently depending on the country's development level.*

### 3. Empirical analysis: Data and sample and methodology

### 3.1. Data and Sample

*3.1.1. Dependent variable*

This paper aims to explore how the perception of corruption perception of a country could influence intrapreneurial activities, focusing on a sample of 92 countries from 2012 to 2019. Data was collected as follows: firstly, we gather a dataset coming from the Global Entrepreneurship Monitor (GEM), which is a multinational survey conducted since 1999, with an increasing number of countries involved, coming from the regions as such classified by GEM of Middle East and Africa, Asia and Pacific, Latin America and Caribbean, and Europe and North America. This survey intends to collect as many forms of entrepreneurship as possible.

As commented before, we focus on Employee Entrepreneurial Activity (our variable eea), that is, the % of adults employed for established companies which are currently implementing a business idea, either a new good/service or business unit for their employer) (see Annex I for a detailed explanation of the dependent and the explanatory variables). With respect to our dependent variable, the GEM reports clarifies that "Although entrepreneurship is often seen as a solitary activity, in practice much entrepreneurial activity is conducted with, and for, others. One example of this is the entrepreneurial employee ("intrapreneur"), who identifies, develops and pursues new business activities as part of their job. The GEM APS asks whether individuals are developing new activities for their employer, such as developing or launching new goods or services, or setting up a new business unit".

We will also show information about other entrepreneurship-related variables such as Nascent Entrepreneurship Activity or *ner*: % of adults starting a new business in the last 3 months, *nbor* or New Business Owner Activity, which is the % of adults who have started a new business more than 3 months ago but for less than 42 months; *tea* is Total early-stage Entrepreneurial Activity and is the sum of *ner + nbor*. This tea variable is usually the most commonly used variable by articles exploring entrepreneurship and using the GEM database. We also include the Entrepreneurial Intentions (*ei*) too, that is, the percentage of working adults who state they intend to start a business in the next three years. All the variables described in this last paragraph will not be used in the main analysis but are explored in the descriptive statistics table for the sake of information and because they are useful to understand the methodology too.

*3.1.2. Independent and Control variables*

With respect to corruption, we use the Corruption Perceptions Index, as elaborated by International Transparency, which is a German non-governmental organization founded in 1993. Its most notable publications are the Global Corruption Barometer and the Corruption Perceptions Index –CPI-. Because this last one is computed as an index "which ranks countries by their perceived levels of public sector corruption, as determined by expert assessments and opinion surveys" we think it's the proxy that better fits our work's explanatory variable of interest. The CPI defines corruption as "the misuse of public power for private benefit" as is elaborated annually. For constructing the explanatory variable in a more intuitive way, we have calculated the variable in a reverse way so that higher values correspond to a higher perception of corruption.

As Alvarez et al. (2011), we use the GDP per capita growth as a control variable.

### 3.2. Methodology

The methodology of this work is not simple. Our database is a panel data, with a limited t and a large N. When looking at the descriptive statistics of our data, the first thing that caught our attention was the inertia that all the entrepreneurship-related variables showed within each country. France for example always shows a value of about 5 for the tea variable for the whole period, Iran exhibits a value of approximately 6 for the ner variable, Saudi Arabia shows a value of about 30 for the ei variable for all the years. [2] The fact that these dependent variables (either ei, ner or eea) show little variability create serious econometric issues and make the methodology heavily dependent on this fact.

In table 1 we can have a look at the reported variation coefficients of the dependent variable along with the other aforementioned entrepreneurship-related variables, which have been calculated within each observation (country) and then taken as a mean for the whole sample. The variation coefficient demonstrates this low variability. This has severe implications: for example when we added the dependent variable lagged one period, and even when we add more periods, its coefficients became highly significant, but because the best predictor for the eea variable in, for example, Spain, with a period mean of 1.738, is the lagged value of the same variable.

---

2 Results not reported

The features of this data do not make the case for the use of a GMM model formulation, neither from a theoretical point of view, as it is not a partial adjustment or an adaptative expectations model, nor from an statistical point of view: when we run a LSDV regression for every dependent variable (thus adding individual dummies for each observation) the significance of the lagged dependent variable is completely lost. The addition of the individual dummies are all significant taking into account a F-test of nested models, which shows very high values. [3] Actually, when running regressions for our dependent variable exclusively on the individual dummies and without any other explanatory variable, the R2 are surprisingly high, running from 58 to 79%, concluding that the individuality explains much of the variance of the dependent variable. [4] What this evidence is showing is that the influence of each national environment, which involves legal, social and institutional factors are crucial, and all these account for each strong individual effect, as these variables are very inertial and only move in the very long run. This leaves not too much space for the influence of other variables, and the aim of this work is to check whether despite this fact, the corruption could still have its part of impact on the different measures of entrepreneurship.

Table 1. Descriptive statistics for our dependent variable (and other entrepreneurship-related variables) and the explanatory and control variables.

|        | Observ. | Mean  | *Mean 1* | *Mean 2* | *Mean 3* | St. dev. | Min.   | Max.  | Var. coef. |
|--------|---------|-------|----------|----------|----------|----------|--------|-------|------------|
| eea    | 395     | 3.19  | *1.23*   | *2.04*   | *5.22*   | 2.60     | 0.00   | 12.62 | 0.38       |
| ner    | 466     | 7.40  | *11.17*  | *8.22*   | *5.36*   | 5.05     | 0.80   | 27.00 | 0.26       |
| nbor   | 466     | 5.80  | *11.24*  | *6.38*   | *3.52*   | 4.32     | 0.17   | 28.13 | 0.22       |
| tea    | 466     | 12.91 | *21.79*  | *14.27*  | *8.73*   | 7.82     | 2.10   | 41.00 | 0.20       |
| ei     | 465     | 22.68 | *41.08*  | *26.46*  | *12.84*  | 15.92    | 2.00   | 79.80 | 0.21       |
| corrupt| 460     | 46.87 | *65.71*  | *57.02*  | *28.99*  | 19.14    | 8.00   | 85.00 | -          |
| gdpgr  | 466     | 1.74  | *2.22*   | *2.02*   | *1.25*   | 2.76     | -14.07 | 23.99 | -          |

---

3 Results not reported

4 All these results are available upon request

We also include the mean subsampled by region (means of the 1, 2 and 3 regions relatively). The number of observations corresponds to the whole sample.

As we can see in the descriptive statistics, eea is quite important in the developed countries (region 3 in our database). For example for the 2017 GEM database the highest EEA rates corresponded to North America 7.9% and Europe 4.4%; Asia and Oceania exhibit a value of 3.1%, and the lowest eea rates are seen in Africa 0.9% and latin American and Caribbean countries (LAC): 1.6%

As we are dealing with panel data, our approach will be to use the Within-Groups estimation (WG) or the Generalized Least Squares (GLS) depending on the result of the Hausman test. These two approaches make it possible to control for this strong individual effects, either by making the results more consistent (WG) or more efficient (GLS). Another issue some authors have raised is that the impact of corruption may not be linear in the regressors. A quick glance to the residuals of the OLS [5] confirms that this could potentially be the case, and thus we will also run the regressions with the square of the explanatory variable.

## 4. Results and discussion

Table 2 shows the results of the main regression. [6] Multicollinearity has been checked and the value has not been higher than 5 in any regression. [7]

In column 1 results of the simplest analysis show that a higher corruption perception is negatively related to intrapreneurship, confirming the 'sand the wheel' hypothesis, and that this influence is significant. In all regressions of Table 2 we have considered the possibility that the perception of corruption that may influence intrapreneuship in year t is the lagged value of cpi, more than the current value of the said variable. When running this alternative specification for the one and two years lagged perception of corruption for the 3 columns the results are very similar. (results not reported). Column 2 shows the result of the analysis exploring the possibility of a

---

[5] All these results are available upon request

[6] We have also run model 3 with multiplicative dummies for the control variable gdpgr but results showed these dummies were not significant and we have not finally included them.

[7] Multicollinearity is checked between eea, cpi and gdpgr and the maximum value was 1.12. We exclude cpi2 from this analysis, as multicollinearity has to be solved when doing an estimation of the independent effect of two variables which happen to be correlated by chance. This is not the case here as by construction, when cpi changes, by force cpi2 is going to change too.

quadratic relation. The evidence confirms that there is a quadratic relation and that corruption initially hampers intrapreneurship till a certain point but then it exerts a positive influence in a 'grease the wheel' style,

The GEM database classifies the countries by development, splitting the sample into three major regions, where Region 1 includes the least developed and 3 the more developed countries. In order to check whether the different level of development is influenced differently by the perception of corruption, we implement this analysis in column 3 by adding multiplicative dummies for the explanatory variable and its squared form (cpi2). Region 2 is used as the dummy base as it is the most numerous group.

**Table 2. Regressions of the intrapreneurship variable on the perception of corruption.**

| Models | I | | II | | III | |
|---|---|---|---|---|---|---|
| | Coefficient (z or t-value) | Sig. | Coefficient (z or t-value) | Sig. | Coefficient (z or t-value) | Sig. |
| cpi | -0.099 -(11.930) | *** | -0.245 -(6.420) | *** | -0.196 -(4.400) | *** |
| Mult.dummy cpi*region1 | | | | | -0.092 -(1.930) | * |
| Mult.dummy cpi*region3 | | | | | 0.074 (1.840) | * |
| cpi2 | | | 0.002 (3.890) | *** | 0.001 (2.450) | ** |
| Mult.dummy cpi2*region1 | | | | | 0.001 (1.900) | * |
| Mult.dummy cpi2*region3 | | | | | -0.001 -(1.490) | |
| gdpgr | -0.014 -(0.490) | | -0.001 -(0.040) | | 0.004 (0.140) | |
| Constant | 7.816 (17.790) | *** | 10.484 (13.260) | *** | 8.851 (8.740) | *** |
| Hausman Test | 3.06 | | 1.09 | | 3.99 | |
| Wald χ2 (GLS) or F (WG) | 142.28 | *** | 179.90 | *** | 209.77 | *** |
| $R^2$ | 59.05 | | 59.74 | | 63.11 | |

Dependent variable is eea; cpi is = 100 - reported Corruption Perceptions Index, as reported by International Transparency, gdpgr is growth of gdp per capita, cpi2 is the square of cpi, Mult.dummy cpi*region1 is the multiplicative dummy that takes the value of cpi when the region is 1, and 0 otherwise. The rest of the multiplicative dummies are calculated likewise. Region 2 is the dummy base, as is the larger subsample. *, **, and *** indicate statistical significance at 10%, 5%, and 1%, respectively.

Results of column 3 confirm hypothesis 2 and show that while the corruption continues to exert a negative impact on intrapreneurship, more obvious in the least developed countries, for the more developed countries this effect is weaker (the sum of the variable and the multiplicative dummy is still negative, but smaller than for the other 2 regions). Corruption in its quadratic form continues to be positive, but again, its effect is stronger for the least developed countries and softer for countries in Region2, and inexistent for the more developed ones, as the multiplicative dummy is not significant.

The growth of GDP is not significant in any regression, the $R^2$ are quite high, and the Hausman test show in every case that the suitable analysis is the Generalized Least Squares estimation, corresponding to a random effects (or more precisely error component model specification).

## 5. Conclusions

Intrapreneurship can serve as a magnificient growth engine for many countries, as it implies innovation in new products, services or business units. In the last decades there have been a number of studies that have aimed to analyze the impact of corruption on entrepreneurship. We use the GEM database to use the variable *Employee Entrepreneurial Activity* as our dependent variable, representing intrapreneurship. To the best of our knowledge this is the first time that a work explores the impact of corruption on intrapreneurship. Understanding this relation is important as intrapreneurship is vital in many countries, and because the relation can be different from that of corruption with entrepreneurship. The perception of corruption may be viewed differently from a wannabe entrepreneur than from someone who is already working for a company, as he/she may have a more experienced and realistic vision.

Results of our analyses confirm that corruption matters and show that the perception of corruption tends to 'sand the wheel' hypothesis and thus corruption is negative for intrapreneurship. There is a confirmation of a quadratic relationship between corruption and intrapreneurship for the nations not belonging to the more developed countries group. Results also confirm that this relation is different across countries according to the different development level. These results are important, because the negative influence of corruption is particularly damaging for the least developed countries. The impact of corruption on intrapreneur behavior can be an attractive issue for scholars considering that we live in societies sometimes marked by abuses of power and public scandals. Our findings can have significant policy implications for the political and economic authorities, as it emphasizes the role of policymakers in legislating against corruption, overall in the less developed countries, because corruption worsens intrapreneurship overall in these countries, but also for the more developed countries because intrapreneurship is relatively more important for this class of nations.

With respect to the limitations of our research and future lines of investigation it would be useful to disentangle the perception of corruption and detail more which activities includes, as it is probably not the same the impact of a bribe that a company is forced to pay than that of a payment to an official to speed up bureaucratic red tape.
.

ANNEX I. Variables.

| | VARIABLES OF THE MAIN ANALYSIS AND REST OF THE ENTREPRENEURHIP-RELATED VARIABLES. |
|---|---|
| | DEPENDENT VARIABLE |
| eea | Employee Entrepreneurial Activity (EEA, is (adults employed for established companies which are currently implementing a business idea, either a new good/service or business unit for their employer)/ (working age adult population). |
| | EXPLANATORY AND CONTROL VARIABLES |
| cpi | Is the Corruption Perceptions Index, as reported by International Transparency (ranging from 1 to 100) modified so as the higher the values the more perception of corruption (cpi = 100 - reported Corruption Perceptions Index) |
| gdpgr | GDP per capita growth for year t |
| | OTHER ENTREPRENEURHIP-RELATED VARIABLES. |
| ner | Nascent Entrepeneurship Activity: (Adults starting a new business but who have not yet paid salaries, or any other payments, including to the founder[s], for three months or more)/(working age adult population). |
| nbor | New Business Owner Activity: (those already running a new business (who have paid wages, or other payments, including to the founder[s], for three months or more but for less than 42 months)/(working age adult population). |
| tea | Total early-stage Entrepreneurial Activity, is the sum of the former two: ner + nbor, and it is considered to be the most important variable representing entrepreneurship coming from the GEM database. |
| ei | Entrepreneurial intentions represent the percentage of working adults (ages 18-64) who state they intend to start a business in the next three years. |